\documentclass[nofootinbib,twocolumn,preprintnumbers,amsmath,amssymb]{revtex4}
\usepackage{epsfig}

\begin{document}
\title{Constraints on $f(R_{ijkl}R^{ijkl})$ gravity: An evidence against\\ the covariant resolution of the Pioneer anomaly}
\preprint{IPM/P-2007/056}
\preprint{arXiv:0708.0662}

\author{Qasem \surname{Exirifard}}
\affiliation{Institute for Studies in Theoretical Physics and Mathematics (IPM), P.O.Box 19395-5531, Tehran, Iran}

\email{exir@theory.ipm.ac.ir}

\begin{abstract}
We consider corrections in the form of $\Delta L(R_{ijkl}R^{ijkl})$ to the Einstein-Hilbert Lagrangian. Then we compute the corrections to the Schwarszchild geometry due to the inclusion of this general term to the Lagrangian. We show that $\Delta L_3=\alpha_{\frac{1}{3}}(R_{ijkl}R^{ijkl})^{\frac{1}{3}}$ gives rise to a constant anomalous acceleration for objects orbiting the Sun onward the Sun. This leads to the conclusion that $\alpha_{\frac{1}{3}}=(13.91\pm 2.11) \times 10^{-26}(\frac{1}{\text{meters}})^{\frac{2}{3}}$  would have covariantly resolved the Pioneer anomaly if this value of $\alpha_{\frac{1}{3}}$ had not contradicted other observations.

We notice that the experimental bounds on $\Delta L_3$ grows stronger in case we examine the deformation of the space-time geometry around objects lighter than the Sun. We therefore use the high precision measurements around the Earth (LAGEOS and LLR) and obtain a very strong constraint on the corrections in the form of $\Delta L(R_{ijkl}R^{ijkl})$ and in particular $\Delta L=\alpha_n(R_{ijkl}R^{ijkl})^n$. This bound requires $\alpha_{\frac{1}{3}}\leq6.12\times 10^{-29}(\frac{1}{\text{meters}})^{\frac{2}{3}}$. Therefore it refutes the  covariant resolution of the Pioneer anomaly.

\end{abstract}
%\pacs{04.20.Fy, 04.50.+h, 04.60.-m}

\maketitle\section{Dynamics of the empty space-time geometry}
As we currently understand nature matters\footnote{Non-vanishing energy momentum tensor.} deform their surrounding space-time geometry and gravity is a side-effect of this deformation. 

The simplest secular action capable of describing the deformation around matters' distribution, or equivalently the dynamics of the space-time, is the Einstein-Hilbert action:
\begin{equation}\label{EH-action-CQG}
 S_{EH}\,=\,\int d^4x \sqrt{-\det g} R,
\end{equation}
which happens to provide a very good phenomenological description of the gravitational phenomena within the solar system. Let us describe the solar system itself in a covariant language. In so doing we notice that the space-time geometry within the solar system possesses the following covariant characters:
\begin{enumerate}
 \item An almost vanishing Ricci scalar and tensor outside the world lines of the Sun and Planets, $R = 0$, $R_{\mu\nu} = 0$.
\item $R_{\mu\nu\lambda\eta} R^{\mu\nu\lambda\eta} \geq \frac{7.53 \times 10^{-71}}{(\text{meters})^4}$.
\end{enumerate}
It should be highlighted that the first of the above assigned properties to the Solar system is not exact due to the cosmological constant. The presence of a cosmological constant of order $\Lambda\approx 10^{-120} \frac{1}{l_{\text{Planck}^2}}$ seems to be the most economical description of the expansion of the universe which has been indicated by the distance-redshift relation.  Ignoring the cosmological constant, however, seems to be a legitimate approximation when $R_{ijkl} R^{ijkl} \gg \Lambda^2= 10^{-104}\frac{1}{\text{meters}^4}$ holds. Since $R_{ijkl} R^{ijkl}\gg \Lambda^2$ is violated beyond $10^7 AU$ from the Sun, a Ricci-flat geometry approximation to the geometry in the Solar system should be indeed very accurate.

Note that the Newtonian gravitational interaction is the leading term in the effective gravitational interaction assigned to \eqref{EH-action-CQG}. If we had been interested only in the effective gravitational interaction  then we could have simply introduced some desired distance-dependent terms into the phenomenological  effective gravitational potential. ``Distance'', however, is not a covariant quantity. We adhere to the standpoint that allows only generally covariant modification to  \eqref{EH-action-CQG}. Once we add a modification, we can address what the modified term implies for the geometry around the Sun, Earth and even for time-dependent solutions.\footnote{This standpoint also allows us to address what observed history of the  cosmological evolution of the universe implies on the time-independent solutions. This suggests that what causes the cosmological inflationary paradigm provides  a lower bound on the mass of black holes \cite{Exirifard:2008iy}. } This perhaps enables us to employ the high precision measurements around the Earth to test the validity of a covariant correction that is suggested by some accurate observations around the Sun.

From a phenomenological standpoint, the Einstein-Hilbert action is only a model compatible with observational quantities and experimental data. Unborn or approaching finer observations and preciser experiments  may lead to some corrections to the Einstein-Hilbert action. These corrections can be any scalar constructed from the Riemann tensor and its covariant derivatives.

From the theoretical standpoint, the simplest example of the corrections to the action we may consider is an arbitrary functional of the Ricci scalar and Ricci tensor:
\begin{subequations}\label{RicciCorrections}
 \begin{align}
 L &= R + \epsilon \Delta L + O(\epsilon^2),\\
\Delta L &= f(R, R_{ij})\,,
 \end{align}
\end{subequations}
where $\epsilon$ encodes the perturbative expansion. The Lagrangian given by \eqref{RicciCorrections} leads to the following equations of motion
\small
\begin{eqnarray}\label{WTF}
 R_{\mu\nu}+ \epsilon k g_{\mu\nu} f(R,R_{ij})+ \epsilon O_{\mu\nu}[\frac{\partial f(R,R_{ij})}{\partial R},\frac{\partial f(R,R_{ij})}{\partial R_{kl}}] \nonumber \\ \qquad=  O(\epsilon^2),
\end{eqnarray}
\normalsize
where $O_{\mu\nu}$ is a linear operator acting on its arguments, and $k$ is a constant number. We should solve \eqref{WTF} in a perturbative fashion:
\begin{eqnarray}\label{WTFG}
g_{\mu\nu} = g_{\mu\nu}^{(0)} + \epsilon g^{(1)}_{\mu\nu} + O(\epsilon^2)\,,
\end{eqnarray}
where $g_{\mu\nu}^{(0)}$ is a Ricci flat metric. Inserting \eqref{WTFG} in \eqref{WTF} leads to
\begin{eqnarray}\label{WTFE}
\square^{(0)}(g_{\mu\nu}^{(1)}) + k f(R,R_{ij})|_{R_{ij}=0} g_{\mu\nu}^{(0)}+\nonumber\\
+ O_{\mu\nu}[\frac{\partial f(R,R_{ij})}{\partial R}|_{R_{ij}=0},\frac{\partial f(R,R_{ij})}{\partial R_{kl}}|_{R_{ij}=0}]\,=\,0\,,
\end{eqnarray}
where $\square^{(0)}$ is a linear second order differential operator. When $f(R,R_{ij})$ has an expansion in term of its variable around $R_{ij}=0$ then its partial derivative with respect to $R$ or $R_{ij}$ either vanishes or diverges at $R_{ij}=0$.\footnote{Examples: The partial derivatives of $f(R)=R^2$ vanishes at $R=0$ while the partial derivative of $f(R)=R^{\frac{1}{3}}$ diverges.}  If the partial derivatives vanish then a perturbative solution exists. For such case since $O_{\mu\nu}$ is a linear operator then \eqref{WTFE} simplifies to
\begin{eqnarray}
 \square^{(0)}(g_{\mu\nu}^{(1)}) + k f(0,0)\,g_{\mu\nu}^{(0)}\,=\,0\,.
\end{eqnarray}
Note that the above equation could have been obtained from the variation of the Einstein-Hilbert action in the presence of a tiny cosmological constant
\begin{eqnarray}\label{EHL}
 S = \int d^4x \sqrt{-\det g } (R+ \epsilon \Lambda) + O(\epsilon^2)\,,
\end{eqnarray}
where $\Lambda\propto f(0,0)$. The exact solutions  of \eqref{EHL} are known.\footnote{Besides the Solar and cosmological constraints on $f(R)$ or $R^n$ make them not attractive \cite{solar-constraint}. Also look at the appendix A for a discussion on a misunderstanding of $\frac{1}{R}$ gravity.}  So the inclusion of the corrections in the form of a functional of the Ricci scalar and the Ricci tensor does not give rise  to not-yet investigated perturbative corrections to the space-time geometry around the Sun. The same conclusion holds for any functional of the Ricci scalar, Ricci tensor and their covariant derivatives provided the functional has an expansion in terms of its variables around a Ricci flat geometry.

A correction to the action thus would `non-trivially' perturb the space-time geometry around a Ricci flat geometry in case the correction involves the Riemann tensor per se. The simplest of such corrections is perhaps an arbitrary functional of the Riemann tensor squared:
\begin{subequations}
\begin{align}
 L & = R + \epsilon \Delta L + O(\epsilon^2),\\
\Delta L & = L(R_{ijkl}R^{ijkl})\,.
\end{align}
\end{subequations}
Examples of this form of correction include:
\begin{subequations}
 \begin{align}
  \Delta L_1 &= \alpha_1 \,R_{ijkl} R^{ijkl}\,,\\
  \Delta L_2 &= \alpha_{\frac{1}{2}} (R_{ijkl} R^{ijkl})^{\frac{1}{2}}\,,\\
  \Delta L_3 &= \alpha_{\frac{1}{3}} (R_{ijkl} R^{ijkl})^{\frac{1}{3}}\,.
 \end{align}
\end{subequations}
Having modified the dynamics of the space-time in small distances, the Heterotic and type I string theories leads to corrections to the Einstein-Hilbert action which includes $\Delta L_1$. \footnote{In the compactification of type II string theories, we also get non-perturbative world-sheet corrections in the form of $L_1$.}  Note that $\alpha_{\frac{1}{2}}$ is a dimensionless parameter, a constant number. Thus $\alpha_{\frac{1}{2}}$ represents a possible  structure constant for the space-time geometry. On the other hand, the dimension of $\alpha_{\frac{1}{3}}$ is negative with respect to the dimension of $\alpha_1$. $\Delta L_3$ thus may require a modification of the dynamics of the space-time in very small Riemann curvatures.

This work aims to identify the best experimental limits and observational bounds on $\Delta L(R_{ijkl}R^{ijkl})$ and especially $\Delta L=\alpha_n(R_{ijkl}R^{ijkl})^n$ regardless of what theory governs the dynamics of the space-time in a very high or low but not-yet achieved curvature. The work will be organized in the following order:

In the second section, we will add a general correction in the form of  $\Delta L(R_{ijkl}R^{ijkl})$ to the Einstein-Hilbert Lagrangian. After that we will compute the corrections to the Schwarszchild geometry  due to the inclusion of this general term in the Lagrangian.

In the third section, we will obtain the effective modified Newtonian gravitational potential for space-craft in the spherical and static extrema of $R+\epsilon \Delta L + O(\epsilon^2)$.  We show that $\Delta L=\alpha_{\frac{1}{3}}(R_{ijkl}R^{ijkl})^{\frac{1}{3}}$ culminates in a constant anomalous acceleration for the objects orbiting the Sun toward the Sun. This leads us to the conclusion that $\alpha_{\frac{1}{3}}=(13.91\pm 2.11) \times 10^{-26}(\frac{1}{\text{meters}})^{\frac{2}{3}}$ would have covariantly resolved the Pioneer anomaly if this value of $\alpha_{\frac{1}{3}}$ had not contradicted other observations. We notice that  the experimental bounds for such correction grows stronger for the space-time geometry around objects lighter than the Sun.

In the fourth section, we will utilize the high precision measurements around the earth (LAGEOS and LLR) to obtain a  strong limit on the corrections in the form of $\Delta L=\alpha_n(R_{ijkl}R^{ijkl})^n$.  This bound requires $\alpha_{\frac{1}{3}}$ to be smaller than $6.12\times10^{-29}(\frac{1}{\text{meters}})^{\frac{2}{3}}$, therefore, it clearly refutes the covariant resolution of the Pioneer anomaly.

In the fifth section, we will note that $\Delta L_2=\alpha_{\frac{1}{2}}(R_{ijkl}R^{ijkl})^{\frac{1}{2}}$ gives rise to an effective logarithmic gravitational potential. We will discuss whether $\Delta L_2=\alpha_{\frac{1}{2}}(R_{ijkl}R^{ijkl})^{\frac{1}{2}}$ can describe the anomalous flat rotational velocity curves of the spiral galaxies. We show that a simple correction in the form of $\Delta L_2=\alpha_{\frac{1}{2}}(R_{ijkl}R^{ijkl})^{\frac{1}{2}}$ is not either in agreement with the high precision measurements around the Earth, or fails to describe the flat rotational velocity curves of the spiral galaxies.

In the last section, we will provide a summary of the results.

\section{Generic but simplest corrections around the Sun}
The Schwarszchild metric is an isotropic and static solution to the Einstein-Hilbert action. In four dimensions, in the standard preferred coordinates, it reads
\small
\begin{equation}\label{Schwarzschild}
ds^2 \,= \, - c^2 (1-\frac{\text{r}_h}{r}) dt^2 + \frac{dr^2}{1-\frac{\text{r}_h}{r}} + r^2 (d\theta^2 + \sin^2 \theta d\xi^2)\,,
\end{equation}
\normalsize
where $ \text{r}_h=\frac{2 G\, m}{c^2}$ in which $G$ is the Newton constant, $m$ represents the mass and $c$ stands for the speed of light. We are interested in the space-time geometry around the Sun, we thus set $m= M_{\odot} = 1.98 \times 10^{30}\text{kg}$ for which  $\text{r}_h \equiv \text{r}_{\odot}\sim 3\text{km}$.

As argued in the first section, since the Einstein tensor vanishes for the  Schwarszchild metric, the geometry around the Sun receives corrections in case the corrections to the action involve the Riemann tensor \textit{per se}. We consider the simplest form of these corrections, the ones which are generic functional of the Riemann tensor's square:
\small
\begin{subequations} \label{action}
\begin{align}
&S\equiv \int d^4x \, \boldsymbol{L}[g_{\mu\nu},R_{\mu\nu\eta\gamma}] + O(\epsilon^2),&\\
\label{BoldL}
&\boldsymbol{L}[g_{\mu\nu},R_{\mu\nu\eta\gamma}]= \sqrt{-\det g}\,(R \, +\, \epsilon\, {\cal L}[R_{\mu\nu\eta\gamma} R^{\mu\nu\eta\gamma}]),&
\end{align}
\end{subequations}
\normalsize
where $\epsilon$ is the parameter of the expansion and ${\cal L}$ is a generic functional. Computing the first variation of  \eqref{action} with respect to the metric, we obtain \cite{Wald2}
\small
\begin{equation}\label{WaldEquation}
0 = -\frac{\partial \boldsymbol{L}}{\partial g_{i j}}\,-\,\frac{\partial \boldsymbol{L}}{\partial R_{i \alpha\beta\gamma}} R^{j}_{\ \alpha\beta\gamma}
- 2 \nabla_{\alpha}\nabla_{\beta} \frac{\partial \boldsymbol{L}}{\partial R_{i\alpha\beta j}}\,,
\end{equation}
\normalsize
where partial derivatives are taken assuming that $g_{\mu\nu}$ and $R_{\mu\nu\eta\gamma}$ are independent variables, and the partial derivative coefficients appearing in \eqref{WaldEquation} are uniquely fixed to have precisely the same tensor symmetries as the varied quantities. Note that our conventions are such that $R_{\phi\theta\phi\theta}$ as well as the Ricci curvature scalar are positive for the standard metric on the two-sphere. We then insert the explicit form of $ \boldsymbol{L}$ presented in \eqref{BoldL} into \eqref{WaldEquation} to obtain
\small
\begin{subequations}\label{Wald-Expanded}
\begin{align}\label{Wald-Expanded-1}
&Eq_{ij}\equiv \text{r.h.s. of \eqref{WaldEquation}}=Eq^{(0)}_{ij} + \epsilon Eq^{(1)}_{ij} + O(\epsilon^2)\,,\\
&Eq^{(0)}_{ij}=R_{ij} - \frac{1}{2} g_{ij} R\,,\\
\label{Wald-Expanded-2}
&Eq^{(1)}_{ij}=2 R^{i}_{\ \alpha\beta\gamma} R^{j\alpha\beta\gamma}
\frac{\partial {\cal L}[x]}{\partial x}|_{_{x={\cal R}^2}}
- \frac{1}{2} g_{ij} {\cal L}({\cal R}^2)\\\nonumber
& \ \ \ \ \ \ \ \ ~- 4 \nabla^{\alpha}\nabla^{\beta}(\frac{\partial {\cal L}[x]}{\partial x}|_{_{x={\cal R}^2}} R_{i\alpha\beta j})\,,
\end{align}
\end{subequations}
\normalsize
whereafter ${\cal R}^2\equiv R_{\mu\nu\eta\gamma}R^{\mu\nu\eta\gamma}$ is inferred.

In the following lines from the outset, we consider the perturbations around the Sun in the standard preferred coordinate:
\small
\begin{subequations}\label{sch}
\begin{align}\label{sch-1}
&ds^2 =  - A(r)\, c^2\, dt^2 + B(r){dr^2} + r^2 (d\theta^2 + \sin^2 \theta d\phi^2)\,,&
\\
\label{sch-2}
&A(r)  = (1-\frac{\text{r}_{\odot}}{r}) (1+ \epsilon \, a(r) )\,+\,O(\epsilon^2)\,,&
\\
\label{sch-3}
&B(r) = \frac{1}{A(r)}(1+ \epsilon \,b(r))\,+\,O(\epsilon^2)\,.&
\end{align}
\end{subequations}
\normalsize
Now let us utilize
\small
\begin{equation}\label{R2}
{\cal R}^2\simeq\frac{12\,\text{r}_h^2}{r^6} + \cdots
\end{equation}
\normalsize
and define
\small
\begin{eqnarray}\label{define-L(r)}
L(r) &\equiv& {\cal L}[{\cal R}^2],\\
\tilde{L}(r)&\equiv&\frac{\partial {\cal L}[x]}{\partial x}|_{_{x={\cal R}^2}}\,=\, -\frac{r^7}{72\,\text{r}_{\odot}^2}\, L'(r)\,+\,O(\epsilon).
\end{eqnarray}
\normalsize
It follows that the non-vanishing and independent components of \eqref{Wald-Expanded} for \eqref{sch} are
\small
\begin{subequations}\label{Eq0-Sch}
\begin{align}
&\frac{1}{c^2}Eq_{tt}^{(0)}\,=\, \frac{\epsilon}{r^2}(1-\frac{\text{r}_{\odot}}{r})((r-\text{r}_{\odot})(a'+b')-a +b)\,+\,O(\epsilon^2)\,,& \\
&Eq_{rr}^{(0)}\,=\,\frac{\epsilon}{r(r-\text{r}_{\odot})}((r-\text{r}_{\odot})\, a'\,+\,a - b)\,+\,O(\epsilon^2)\,,&\\&Eq_{\theta\theta}^{(0)}\,=\,\frac{\epsilon r}{2}((r-\text{r}_{\odot}) a'' + 2 a' - b' + \frac{b'\, \text{r}_{\odot}}{2 r})\,+\,O(\epsilon^2)\,,&
\end{align}\end{subequations}\normalsize
and
\small
\begin{eqnarray}\label{Eq1-Sch}
\frac{Eq_{tt}^{(1)}}{c^2({r-\,\text{r}_{\odot}})}&\,=\,&
-\frac{2\text{r}_{\odot}}{r^6}(2 r (r-\text{r}_{\odot}) \tilde{L}''
- (2 r - 3 \text{r}_{\odot}) \tilde{L}')\, \nonumber\\
&&+\frac{r L' + 6 L}{12\,r} \,\,+\,O(\epsilon^2)\,,
\\ \nonumber
(r-\,\text{r}_{\odot})Eq_{rr}^{(1)}&=&-\frac{r^2 L' + 6 r L}{12} \,-\frac{2\,\text{r}_{\odot}\,(- 3\,\text{r}_{\odot} + 2 r)}{r^4}\, \tilde{L}'\,,\\
&&\,+\,O(\epsilon^2)\,
\\
Eq_{\theta\theta}^{(1)}&=&
-\frac{2}{r^3}(r (r - \text{r}_{\odot}) \tilde{L}''- (2r - 3 \, \text{r}_{\odot})\,\tilde{L}')\,,\nonumber
\\ & &-\frac{r L' + 6 L}{12}\, r^2\,+\,O(\epsilon^2)\,,
\end{eqnarray}
where the dependence of the various functions on $r$ is understood. Then \eqref{Wald-Expanded} gives rise to a non-homogeneous second order differential equation for $a(r)$ and a non-homogeneous first order differential equation for $b(r)$.  In accordance with the preceding studies of string world-sheet corrections to various black holes \cite{Callan,martin,Exirifard:2006qv} we demand that the corrections must not diverge on the possible event horizon at $r\,=\,\text{r}_{h}$, provided  the contribution of ${\cal L}({\cal R}^2)$ remains bounded on the horizon. We
 find out that this precondition is satisfied by the following solution  
\small
\begin{subequations}\label{a-b(r)}
\begin{align}
&b(r) \,=\, -4\,\text{r}_{\odot}\, \int_{\text{r}_{\odot}}^{r}dx\,\frac{\tilde{L}''(x)}{x^2}\, , \\
&a(r) \,=\, -\frac{1}{2(r -\,\text{r}_{\odot})} \int_{\text{r}_{\odot}}^{r}\frac{dx}{x^4}\left[
{-12\,\text{r}_{\odot}^2\,(x \tilde{L}'(x) + \tilde{L}(x))+ }\right.\\\nonumber
&\left.\hspace{2.5cm} + 8\, \text{r}_{\odot}\, x^2 \tilde{L}(x) + x^6 L(x) + 2 x^4 b(x)\right],
\end{align}
\end{subequations}
\normalsize
The above boundary conditions also reproduce the correct results  around ordinary stars. In order to illustrate this claim, we note that we are solving the equations of motion in a perturbative fashion. We note that $R_{ijkl}R^{ijkl}$ inside the Sun is larger than  $R_{ijkl}R^{ijkl}$  outside the Sun. Since the perturbation that we are interested in grows as the Riemann curvature decreases (see the end of third section) then the perturbation holds valid inside the Sun.  Let \eqref{WTFG} represent the perturbative solution.  Inserting this perturbative expansion in the equation of motion leads to a second order non-homogeneous differential  equation for $g_{\mu\nu}^{(1)}$. Let us seperately write the equations of motion of $g_{\mu\nu}^{(1)}$ for inside and outside the matter's distribution:
\begin{eqnarray}
\Box^{(0)}.g^{(1)}_{in}\,=\, F[g^{(0)}_{in}]\,,\\
\label{revisedboundary2}
\Box^{(0)}.g^{(1)}_{out}\,=\, F[g^{(0)}_{out}]\,,
\end{eqnarray}
where indices are understood but not written and $F$ stands for the non-homogeneous part. Outside the star we have $g^{(0)}_{out}=g_{\text{Schwarszchild}}$. The consistency of the perturbation requires that $g^{(1)}_{in}$ remains bounded inside the star. Requiring the existence of the metric and its first derivatives on the surface of the star then provides the physical boundary condition for $g^{(1)}_{out}$. This boundary condition, however, requires first solving the equations inside the star.  Therefore a method capable of reproducing the physical boundary condition (which does not   need solving the equations inside the star) is appreciated .
 
Since the equation for $g_{out}^{(1)}$ is linear, $g_{out}^{(1)}$ for a star reads
\begin{equation}
g^{(1)}_{out} = c_1 . g^{(1)}_{div} + c_2 . g^{(1)}_{con} + g^{(1)}_{\text{Non-homogeneous}}
\end{equation}
where $g_{div}^{(1)}$ and $g_{con}^{(1)}$ are solutions of homogeneous part of \eqref{revisedboundary2} that respectively diverges or converges when they are extrapolated toward the Schwarzschild radius associated to the central mass. $c_1$ and $c_2$  should be chosen such that the metric inside and outside the star match each other. In the Einstein-Hilbert action, and for a star that is formed by collapse of dust, $c_1$ and $c_2$ can only depend on the total central mass:  $c_1=c_1(M)$, $c_2=c_2(M)$.  On the other hand $c_1$ and $c_2$ are numbers while $M$ is parameter with a dimension. Therefore $c_1$ and $c_2$ must  be independent of the mass of the star. Subsequently, once a physical criterion fixes $c_1$ and $c_2$ for a spherical mass distribution then they are fixed for every spherical time-independent distribution of mass. Since  $c_1=0, c_2=1$ for a large black hole then it holds    $c_1=0, c_2=1$ for the Sun. We thus conclude that  \eqref{a-b(r)} also describes the perturbation around the Sun.

It is worth noting that though the precondition we employed does not fix all the boundary conditions, it fixes the radial dependent part of $a$ and $b$. The constant parts in $a$ and $b$ do not affect the  force exerted onto a spacecraft, the quantity we need in the next sections. In particular for $\epsilon {\cal L}[{\cal R}^2]=\alpha_n (R_{\mu\nu\lambda\eta}R^{\mu\nu\lambda\eta})^n$, the large radius behavior of $a(r)$ and $b(r)$ simplifies to
\small
\begin{subequations}\label{abn(r)}
\begin{align}
&a(r) \,\simeq\, -4 (12)^n  n(1-n) \alpha_n r_\odot^{2-4n} \frac{(\frac{r}{r_\odot})^{3-6n}-1}{3-6n},\\
%
%\frac{4 (12)^n n (n-1)}{3 (2n-1)} \text{r}_\odot^{2n-1}\, \alpha_n\,r^{3-6n}\\
&b(r) \,\simeq\, -(\frac{5}{2}-3 n) a(r)\, ,
\end{align}
\end{subequations}
\normalsize
where $n\neq\frac{1}{2}$ is presumed.\footnote{It is not legitimate to extrapolate \eqref{abn(r)} toward far infinity since it has been obtained after expanding the equations for regions of the space-time which meets $R_{\mu\nu\lambda\eta} R^{\mu\nu\lambda\eta} \geq \frac{7.53 \times 10^{-71}}{(\text{meters})^4}$.} For $n=\frac{1}{2}$ one finds
\small
\begin{equation}\label{ab-ln(r)}
a(r) \,\simeq\, b(r) \,\simeq\, -({12})^{\frac{1}{2}} \alpha_{\frac{1}{2}} \ln(\frac{r}{r_\odot})\,.
\end{equation}
For sake of simplicity, we will consider only the corrections in the form of $\epsilon {\cal L}[{\cal R}^2]=\alpha_n (R_{\mu\nu\lambda\eta}R^{\mu\nu\lambda\eta})^n$ but the results can be directly generalized to a general case. We also note that the perturbations increase as we go further away from the Sun. We, however, shall address the regime of the validity of the perturbation in the fourth section after having independently obtained the experimental bounds on $\alpha_n$.

\section{Effective potential and the covariant resolution of the Pioneer anomaly}
The effective potential for spacecrafts in the spherical and static geometry of \eqref{sch} is \cite{blau}
\small
\begin{equation}\label{Veff}
V_{eff}(r)\,=\,\frac{l^2}{2 \,r^2\,B(r)}-\frac{E^2}{2\,c^2\, A(r)\, B(r)}+\frac{c^2}{2\,B(r)}\,,
\end{equation}
\normalsize
where the equatorial plane is chosen to be orthogonal to the angular momentum, and $E$ stands for the energy (per unit rest mass) and $l$ represents the magnitude  (per unit rest mass) of the angular momentum.

Inserting the $\epsilon$ perturbative expansion series of $A(r)$ and $B(r)$ into \eqref{Veff} yields
\small
\begin{subequations}\label{Veff-app}
\begin{align}\label{Veff-apps}
&V_{eff}(r)\,=\,V_{eff}^{(0)}(r) + \epsilon V_{eff}^{(1)}(r) + O(\epsilon^2)\,,\\
\label{Veff-app-0}
&V_{eff}^{(0)}(r)\,=\,-\frac{G\, M_{\odot}}{r} + \frac{l^2}{2\,r^2}-\frac{G\,M_{\odot} l^2}{r^3 c^2}\,,
\\
\label{Veff-app-1}
 & V_{eff}^{(1)}(r)\,=\,\frac{E^2-c^4}{2 c^2} b(r) + \frac{c^2}{2} a(r)\,+\,O(\frac{l^2}{2 r^2}),
\end{align}
\end{subequations}
\normalsize
where a constant term is tacitly recognized in \eqref{Veff-app-0}. We  notice that $V_{eff}^{(0)}(r)$ is the effective potential for the Einstein-Hilbert action. Thus it is $\epsilon V_{eff}^{(1)}(r)$ which gives rise to the anomalous accelerations for spacecrafts deployed to explore the outer solar system.

The Pioneers 10 and 11, spacecrafts deployed for exploring the outer solar system are reported to have experienced a constant anomalous acceleration  of magnitude $a_{p}\equiv(8.74\pm 1.33)\times10^{-10} \frac{m}{s^2}$ in the direction toward the Sun at a distance of  $20-70 \text{AU}$ (Astronomical Units) from the Sun\footnote{The major part in the error bar of $a_p$ is systematic. The statistical error is only $0.01\times10^{-10} \frac{m}{s^2}$. } \cite{Nieto:2003rq}. What causes this anomaly might be on-board systematics, but the smoking gun has not been found yet \cite{Anderson:2001ks}. Let us see if a correction in the form of $\Delta L=L(R_{ijkl}R^{ijkl})$ may resolve the Pioneer anomaly.

The effective potential culminating to the observed constant anomalous acceleration of the Pioneers is
\small
\begin{equation}\label{pioneers-effective-potential}
\epsilon V_{eff}^{(1)}(r) \,=\,-\, a_p\, r\,,
\end{equation}
\normalsize
from now on we assume that \eqref{pioneers-effective-potential} is valid for $r \in 0-70$AU.  

The anomalous acceleration is obtained from analyzing the Doppler shift of the electromagnetic wave that the Pioneers had been sending assuming that the space-time geometry around the Sun coincides to the Schwarszchild geometry.  Deviation from the Schwarschild geometry affects the gravitational red/blue shifts and subsequently alters the anomalous acceleration assigned to the Pioneers \cite{Planetory-constraint,Jaekel:2005qz, Jaekel:2006me}.\footnote{I  appreciates the comment made by the referee of the CQG that leads to adding a discussion on this issue.}  We first assume that the deviation from the Schwarszchild geometry needed to describe the Pioneer anomaly does not significantly affect the gravitational red/blue shifts of the signals sent by the spacecrafts. We confirm this assumption after having identified the deviation from the Schwarzschild geometry.  

If the Pioneer effective potential \eqref{pioneers-effective-potential} is due to gravitational effects of the Sun then   \eqref{Veff-app-1} and \eqref{pioneers-effective-potential} results
\small
\begin{equation}\label{ab-equation}
\epsilon\,\left(\frac{E^2-c^4}{2 c^2} b(r) + \frac{c^2}{2} a(r)\right)\,\simeq\,-\, a_p\, r\,.
\end{equation}
\normalsize
One notices that the Pioneers have classical velocity and $\frac{E^2-c^4}{2 c^2}\simeq\frac{v^2}{4}\ll c^2$ whereafter $v$ stands for the radial velocity of the spacecrafts with respect to the Sun. Therefore \eqref{ab-equation} can  be  further approximated to
\small
\begin{equation}\label{appro-a}
\epsilon\,a(r)\,=\,-\,\frac{2 a_p}{c^2}\, r\,+\,O(\frac{v^2}{c^2})\,,
\end{equation}
\normalsize
where rearranging the terms is induced. Note that in these distances it holds $\epsilon a(r)\ll \frac{\text{r}_\odot}{r}$ and the approximation in \eqref{appro-a} is much lesser than the error bar in $a_{p}$.

\begin{figure}\label{Pioneer-Detector}
   \epsfxsize=8cm
   \epsfbox{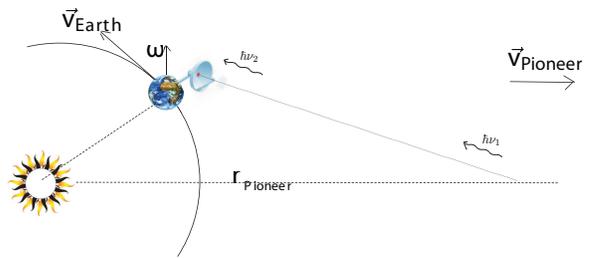}
   \caption{\small \textit{The acceleration of the Pioneer spacecrafts are identified by analyzing the Doppler shift of the spacecrafts' emission. The gravitational red/blue shifts are part of this analyze.  Altering the geometry affects the gravitational red/blue shifts. Therefore, the back reaction of the change of geometry to the  constant anomalous acceleration should be addressed if the anomaly is due to a covariant correction.}}
\end{figure}

Eq. \eqref{appro-a} identifies the deviation from the Schwarzschild geometry. In the following we are going to argue that this deviation does not significantly alters the gravitational Doppler shifts  to which the  anomalous acceleration is assigned. The Pioneer spacecraft\footnote{The Deep Space Network antennas tracked the pioneer spacecrafts with a S-band signal at about 2.11 GHZ. The tracking was done by sending a signal from the Earth which the Pioneers were replying. The simple one-way Doppler shift analyze of this section, however, suffices for our conclusions.} sends a photon at time $t$ in $x=\vec{r}_{\text{Pioneer}}$ of frequency $\nu_1$ in the local frame that is fixed at $\vec{r}_{\text{Pioneer}}$. The detector on the Earth recieves the photon on time $t+T(t)$ with frequency of $v_2$ in the local from that is fixed at $x=\vec{r}_{\text{Detector}}$, see fig. \ref{Pioneer-Detector}.  $T(t)$ can be identified in terms of the initial position of the spacecraft, Earth and the detector.

In the absence of gravity we assign energy of $\hbar \nu_1$ to a photon related to light's wave of frequency $\nu_1$. In the presence of the week gravity, therefore, we should assign energy of $\hbar v + V_{eff}. \frac{\hbar v}{c^2}$ to the total energy for the photon of frequency $\hbar \nu$ where $V_{eff}$ is the effective gravitational potential at where photon is localized.  Requiring the conservation of the energy for the photon sent by the Pioneer yields
\small
\begin{eqnarray}\label{frequency-expansion}
\nu_1 &(1 + \frac{1}{c^2} V_{eff}[r_{\text{Detector}(t+T)}]) =
\nu_2 (1 + \frac{1}{c^2} V_{eff}[r_{\text{Pioneer}(t)}]) \nonumber\\
\to &\frac{\nu_1-\nu_2}{\nu_2} = \frac{1}{c^2} (V_{eff}[r_{\text{Pioneer}(t)}]-V_{eff}[r_{\text{Detector}(t+T)}])\,.
\end{eqnarray}
\normalsize
Employing \eqref{Veff-app} then yields
\small
\begin{eqnarray}
c^2 \frac{\Delta \nu}{\nu} & = & V_{eff}^{(0)}[r_{Detector}(t+T)] - V_{eff}^{(0)}[r_{Pioneer}(t)] + \nonumber\\
&& + \frac{\epsilon}{2} (a[r_{Detector}(t+T)] - a[r_{Pioneer}(t)] )\,.
\end{eqnarray} 
\normalsize
We note that the time that a photon needs to travel from the spacecraft to the detector also has an $\epsilon$ expansion:
\small
\begin{eqnarray}\label{T-expansion}
T &=& T^{(0)} + \epsilon T^{(1)} + O(\epsilon^2),\\
r_{D.}(t+T) &=& r_{D}(t+T^{(0)}) + \epsilon \dot{r}_{D}(t+T^{(0)}) T^{(1)} + O(\epsilon^2)\nonumber\,, 
\end{eqnarray}
\normalsize
where $r_{D.}=r_{\text{Detector}}$. Inserting \eqref{T-expansion} in \ref{frequency-expansion} leads to
\small
\begin{eqnarray}\label{periodic-doppler}
\frac{\Delta \nu}{\nu} &=& (\frac{\Delta \nu}{\nu})_{\text{E.-H.}} + 
\frac{\epsilon}{c^2} \frac{ \partial V_{eff}^{(0)}(r) }{\partial r}|_{r=r_{\text{D.}}} \frac{\partial r_{D.}(t)}{\partial t} T^{(1)} + \nonumber\\
&& + \frac{\epsilon a_p}{c^2}[r_{\text{Pioneer}}(t)-r_{\text{Detector}}(t+T^{(0)})]\,,
\end{eqnarray}
\normalsize
where $(\frac{\Delta \nu}{\nu})_{\text{E.-H.}}$ is the prediction of the Einstein-Hilbert action. This states that ignoring the deviation from the Einstein-Hilbert action leads to a systematic error in $\frac{\Delta \nu}{\nu}$ given by
\small
\begin{eqnarray}
(\frac{\Delta \nu}{\nu})_{missing} &=& 
\frac{\epsilon}{c^2} \frac{ \partial V_{eff}^{(0)}(r) }{\partial r}|_{r=r_{\text{D.}}} \frac{\partial r_{D.}(t)}{\partial t} T^{(1)} + \\
&& + \frac{\epsilon a_p}{c^2}[r_{\text{Pioneer}}(t)-r_{\text{Detector}}(t+T^{(0)})]\,,\nonumber
\end{eqnarray}
\normalsize
which in turn results to a systematic error in determining the acceleration of the spacecraft
\small
\begin{eqnarray}
(a)_{missing} &=& c \frac{d}{dt}(\frac{\Delta \nu}{\nu})_{missing}\,.
\end{eqnarray}
\normalsize
We notice that $T^{(0)} \approx \frac{r_{\text{Pioneer}}}{c}$. $T^{(1)}$ should be proportional to $a_p$. Therefore $T^{(1)}\approx a_p \frac{(T^{(0)})^2}{c}$. This helps us to obtain the order of magnitude of  the terms present in $a_{missing}$:
\small
\begin{equation}
\frac{(a)_{missing}}{a_p} = O(\frac{v_{\text{Earth}} }{c})
 + O(\frac{  v_{ \text{Pioneer} } }{c}) 
+ O(\frac{ R_{\text{Earth}} \omega_{\text{Earth}} }{c})\,.
\end{equation}
\normalsize
Therefore, due to the error bar in $a_{p}=(8.74\pm 1.33)\times10^{-10} \frac{m}{s^2}$, it is legitimate to neglect $a_{missing}$.

We note that the covariant resolution of the Pioneer anomaly gives rise to periodic terms (with periodicity of one day and one year) in the Doppler shift \eqref{periodic-doppler}.  \cite{Levy:2008wz} and \cite{Anderson:2001sg} report that the residual of the fit with constant anomalous acceleration contains clear periodic terms. Ref. \cite{Anderson:2001sg} argues that these periodic terms should be assigned to the Earth and its atmosphere while \cite{Levy:2008wz} discusses that they are somehow fingerprints  of what causes the Pioneer anomaly.  The  periodic terms reported in \cite{Levy:2008wz,Anderson:2001sg}    are at order of $\frac{\Delta \nu}{\nu}\approx \frac{1 mHz}{1 GHz}= 10^{-12}$. The term of periodicity of one year in \eqref{periodic-doppler} is at order of magnitude $\frac{\Delta \nu}{\nu}|_{yr}\approx \frac{a_p}{c^2}{2  AU}= 10^{-15}$, and the term of periodicity of one day in \eqref{periodic-doppler} at order $\frac{\Delta \nu}{\nu}|_{Day}\approx \frac{a_p}{c^2}{ 2 \times 6400 km }= 10^{-19}$. Therefore no covariant resolution of the constant pioneer anomaly    is able to account for the residual periodicity of the fit with constant anomalous acceleration (note that this statement was derived  only from \eqref{appro-a}).

 Now let us come back to the main issue of this section: how does \eqref{appro-a} help us to identify the correction to the action? We see that \eqref{appro-a} beside \eqref{a-b(r)} leads to an integral equation for the correction to the Einstein-Hilbert action. Variation  of this integral equation with respect to $r$ leads to a non-homogeneous linear third order differential equation for $L(r)$. Eq. \eqref{abn(r)} then shows that one solution of this differential equation is
\small
\begin{eqnarray}\label{found-Delta-L}
\epsilon {\cal L}[{\cal R}^2] \,=\, \,\alpha_{\frac{1}{3}}\,\times ({R_{\mu\nu\eta\gamma}R^{\mu\nu\eta\gamma}})^{\frac{1}{3}}\,,
\end{eqnarray}
\normalsize
which leads to
\small
\begin{equation}\label{ab-found}
\epsilon a_{\frac{1}{3}}(r) = -\,\alpha_{\frac{1}{3}}\,(\frac{12}{\text{r}_{\odot}})^{\frac{1}{3}}\,\frac{8 r}{9}\,,~~  b_{\frac{1}{3}}(r) =  \,\frac{3}{2}\,a_{\frac{1}{3}}(r)\,.
\end{equation}
\normalsize
It is worth noting that \eqref{found-Delta-L} is only one solution to the corresponding non-homogeneous linear third order differential equation for $L(r)$. Other solutions differ with \eqref{found-Delta-L} by terms which do not affect the motion of a spacecraft. Since we are interested in a motion of a spacecraft,  we consider only \eqref{found-Delta-L}. Then comparing \eqref{ab-found} to \eqref{appro-a} identifies $\alpha$ to
\small
\begin{equation}\label{alpha0-sun}
\alpha_{\frac{1}{3}}^p \,=\,(13.91 \pm 2.11) \times 10^{-26} (\frac{1}{\text{meter}})^{\frac{2}{3}} \,,
\end{equation}
\normalsize
which would have covariantly resolved the Pioneer anomaly if it had not been in contradiction with the other observations.

\section{Constraints from Earth and Moon}
In the previous section we have examined a general family of the covariant corrections and found the covariant correction capable of describing the Pioneer anomaly. We note that the corrections to the space-time geometry given in eq. \eqref{ab-found} would increase if the mass of the Sun decreases (recall that $r_\odot=\frac{2G M_\odot}{c^2}$). It implies that the accurate measurements of the geometry around the Earth would provide a strong constraint on the covariant corrections in the form of:
\small
\begin{equation}\label{gravity-action-in-vacuum}
S\,=\,\int d^4 x\,\sqrt{-\det g} (R + \alpha_n (R_{\mu\nu\eta\gamma}R^{\mu\nu\eta\gamma})^{n})\,.
\end{equation}
\normalsize
Recalling that  \eqref{gravity-action-in-vacuum} and \eqref{Veff-app-1} hold a perturbative expansion like to that of \eqref{a-b(r)} and (\ref{abn(r)}-\ref{ab-ln(r)}) around the Earth, we find that a probe with a classical velocity experiences an anomalous acceleration of magnitude $a_e(r)$, given by:
\small
\begin{subequations}\label{an-Earth}
 \begin{align}
&a_e(r) \, =\, - a_n\, r^{2-6n}\,,\\\label{an-Earth-2}
&a_n\,=\,  2 (12)^n n (1-n) c^2 \left(\frac{2\, G\, M_{\text{Earth}}}{c^2}\right)^{2n-1} \alpha_n\,,
\end{align}
\end{subequations}
\normalsize
at the distance $r$ from the center of the Earth toward the center of the Earth in case that eq. \eqref{gravity-action-in-vacuum} governs the dynamics of the space-time.
To put it another, a satellite in a circular orbit experiences the following  gravitational field around  the earth
\begin{equation}
 F_G \,=\, \frac{GM_{\text{Earth}}}{r^2} + a_n r^{2-6n} \,.
\end{equation}
where $M_{\text{Earth}}$ is the inertial (effective) mass of the Earth. The effective gravitational mass of the Earth is defined by
\begin{equation}\label{GM-theory}
 GM_{\text{Earth}}^{eff}(r) =  r^2 F_G \,=\,GM_{\text{Earth}} + a_n r^{4-6n}\,.
\end{equation}
In contrast to the Newtonian dynamic, the effective gravitational mass is not a radius-independent quantity. Analyzing the circular orbit of any satellite or the Moon identifies the effective gravitational mass of the Earth within that orbit.

\begin{figure}
   \epsfxsize=8cm
   \epsfbox{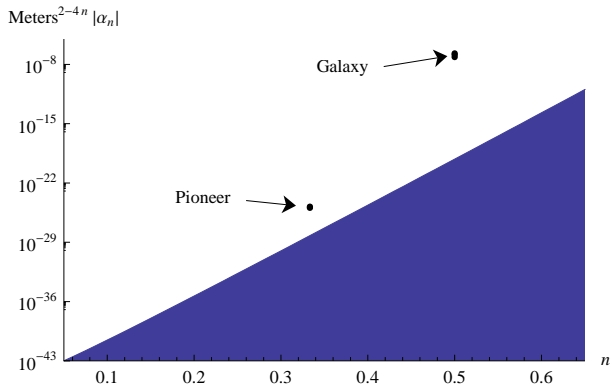}
   \caption{\label{fig1}\small \textit{The blue region is the range of parameters describing the covariant corrections to gravity which are consistent with the accurate measurements of the space-time geometry around the Earth \eqref{an-limit}. The dots represents the values which will be needed to covariantly resolves the Pioneer anomaly or the anomalous velocity curves of the spiral galaxies. The thickness of the dots corresponds to their uncertainties. }}
\end{figure}

The accurate value of the mass ratio of the Sun/(Earth+Moon) from the Lunar Laser Ranging can be combined with the Solar GM and the lunar GM from lunar orbiting spacecrafts \cite{moon-satellites} to give the effective gravitational mass of the Earth in an Earth-centered reference frame with the precision of one part in $10^8$:
\small
\begin{equation}\label{GM-Moon}
GM_{\text{Earth}}^{LLR}(d_{\text{Earth-Moon}})\,=\, 398600.443 \pm 0.004 \frac{\text{km}^3}{s^2}\,,
\end{equation}
\normalsize
where $d_{\text{Earth-Moon}}$ is the distance between the  Moon and Earth \cite{LLR}. The  effective gravitational mass of the Earth has also been measured by various  artificial Earth satellites \cite{satellites}, including the accurate tracking of the LAGEOS satellites orbiting the Earth in nearly circular orbits with semimajor axes about twice the radius of the Earth:
\small
\begin{equation}\label{GM-LAGEOS}
GM_{\text{Earth}}^{LAGEOS}(2r_{\text{Earth}})\,=\,398600.4419\pm 0.0002â€‰\frac{\text{km}^3}{s^2}\,,
\end{equation}
\normalsize
where $r_{\text{Earth}}$ stands for the radius of the Earth \cite{LAGEOS}. We note that within the errors \eqref{GM-LAGEOS} is compatible with \eqref{GM-Moon}.  Recalling \eqref {GM-theory} beside comparing  (\ref{GM-Moon}) to (\ref{GM-LAGEOS}), therefore, leads to
\small
\begin{equation}\label{an-limit-comp}
|a_n| \left((d_{\text{Earth-Moon}})^{4-6n}-(2\,r_{\text{Earth}})^{4-6n}\right) \leq 0.004 \frac{\text{km}^3}{s^2}\,.
\end{equation}
\normalsize
where $|a_n|$ stands for the absolute value of $a_n$. For $n<\frac{2}{3}$, \eqref{an-limit-comp} can be approximated to
\small
\begin{equation}\label{an-limit}
|a_n| (d_{\text{Earth-Moon}})^{4-6n} \leq 0.004 \frac{\text{km}^3}{s^2}\,
\end{equation}
\normalsize
Using \eqref{an-Earth-2}, \eqref{an-limit} results
\begin{equation}\label{limit-alpha-n}
|\alpha_n| \leq \frac{0.004 \frac{\text{km}^3}{s^2}(\frac{G M_{\text{Earth}}}{c^2})^{1-2n}}{2 (12)^n n (1-n) c^2(d_{\text{Earth-Moon}})^{4-6n}}\,,
\end{equation}
The values of $|\alpha_n|$ which meet \eqref{limit-alpha-n} are illustrated in Fig.\ref{fig1} for $0.05\leq n\leq 0.65$.  The limit on $|\alpha_{\frac{1}{3}}|$ is $|\alpha_{\frac{1}{3}}|\leq 6.12\times 10^{-29}(\frac{1}{\text{meters}})^{\frac{2}{3}}$. Therefore $\alpha_{\frac{1}{3}}^p \,=\,(13.91 \pm 2.11) \times 10^{-26} (\frac{1}{\text{meter}})^{\frac{2}{3}}$, which is needed to covariantly resolve the Pioneer anomaly,  is clearly  not compatible with the accurate measurements around the Earth. So the Pioneer anomaly can not be covariantly resolved within the general family of corrections we have considered. This supports the idea that the Pioneer anomaly is on board systematic or due to non-gravitational effects. This idea is in agreement with other independent studies: the precession of the longitudes of perihelia of the solar planets \cite{Lorenzo} or the trajectories of long period comets \cite{comet} have not been reported to experience an anomalous gravitational field toward the Sun of the magnitude capable of describing the Pioneer anomaly.

\begin{figure}
   \epsfxsize=8cm
   \epsfbox{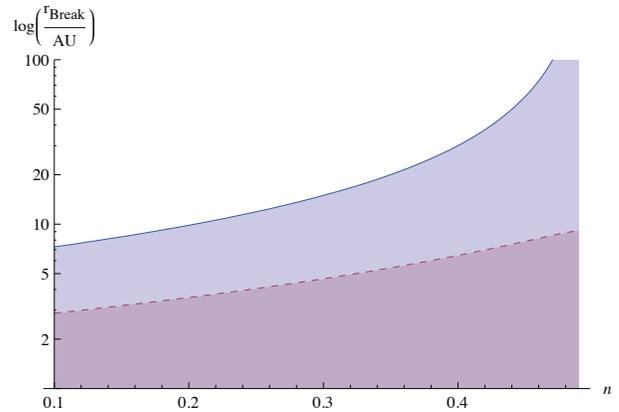}
   \caption{\label{fig3}\small \textit{We have used  the bounds on $\alpha_n$ which is presented in Fig. 1 to find how far from the Sun the perturbation remains valid. The perturbation in terms of $\alpha_n (R^{ijkl}R_{ijkl})$ is valid below the blue continuous line. Below the dashed line, the perturbation around the Schwarszchild geometry is valid.   Note that the y-axis is \textit{both} logarithmic and represents $\log_{_{10}}(\frac{r_{\text{Break}}}{AU})$ where $r_{Break}$ is the minimum distance from the Sun that the `corresponding' perturbation breaks in.}}
\end{figure}

Having obtained the experimental bounds on $\alpha_n$ \eqref{limit-alpha-n}, we would like to find the  minimum distance from the Sun that the perturbation breaks in.\footnote{I thank the comment of the referee of CQG that leads to adding the subsequent paragraphs in this section.} Note that we have assumed that the space-time geometry has a perturbation around the Schwarzschild geometry,
\begin{equation}
-g_{tt} = A(r) = (1- \frac{r_\odot}{r})(1+ \epsilon a(r) + O(\epsilon^2))\,,
\end{equation}
wherein we have assumed that the Schwarszchild geometry describes the space-time geometry with a very good approximation. The existence of the perturbation means that $\epsilon a(r)\ll1$. The Schwarzschild geometry remains a good approximation if $\epsilon a(r)\ll \frac{r_\odot}{r}$. The perturbation breaks when $\epsilon a(r)\approx \frac{r_\odot}{r}$ or $\epsilon a(r)\approx 1$.  Using the bounds on $\alpha_n$ we then obtain
\small
\begin{eqnarray}
\epsilon a(\tilde{r}) \approx 1 \to \frac{\tilde{r}}{r_\odot}\approx(1 + |\frac{3-6n}{12^n 4 n (n-1)}\frac{1}{\alpha_n r_\odot^{2-4n}}|)^{\frac{1}{3-6n}}\nonumber\\
\epsilon a(\tilde{r}) \approx \frac{r_\odot}{r} \to \frac{\tilde{r}}{r_\odot}\approx(|\frac{3-6n}{12^n 4 n (n-1)}\frac{1}{\alpha_n r_\odot^{2-4n}}|)^{\frac{1}{4-6n}}\nonumber\\
\end{eqnarray}
\normalsize
where $\alpha_n$ is such that the bound in \eqref{limit-alpha-n} is saturated, and in the last line it is assumed that $n\neq \frac{1}{2}$. Fig 2. plots $\tilde{r}$. The first observation is that the Perturbation remains valid inside the Solar system. It is also interesting  that the perturbation around the Schwarszchild metric breaks before perturbation in $\alpha_n(R_{ijkl}R^{ijkl})^n$. This means that there exist some regions where in $R+ \epsilon\alpha_n(R_{ijkl}R^{ijkl})^n+O(\epsilon^2)$ is perturbative in the sense that terms of order $\epsilon^2$ can be consistently neglected while the dominant term in $R+ \epsilon\alpha_n(R_{ijkl}R^{ijkl})^n$ is $\alpha_n(R_{ijkl}R^{ijkl})^n$ not $R$. It might be interesting to solve the ``exact" equations in these regions.

We would like to reemphasize  that the combined $LLR$ and LAGEOS measurements provide indeed a strong constraint on the form of the covariant correction. In order to further illustrate the power of this constraint let us investigate if it is satisfied by the covariant corrections proposed in ref. \cite{Saffari:2007zt}.   Ref. \cite{Saffari:2007zt} considers a family of $f(R)$ gravity in the presence of the cosmological constant before showing that in the vicinity of the Sun, there exists a set of $f(R)$ corrections capable of the describing the Pioneer anomaly. It shows that the effective gravitational acceleration is 
\begin{equation} \label{Rahvar}
a_{gravity} = -\frac{G M_\odot}{r^2} - a_{constant}\,,
\end{equation}  
where the second term is ``a constant acceleration [while is] independent of the [considered central] mass". It then sets  $a_{constant}=a_p=(8.73\pm 1.33) \times 10^{-10} \frac{m}{s^2}$. Ref.  \cite{Saffari:2007zt} however \textit{has not} considered the implication of the covariant correction they studied to the space-time geometry around the Earth. \eqref{Rahvar} implies the following effective gravitational acceleration around the Earth:
\begin{equation} 
a_{gravity} = -\frac{G M_{\text{Earth}}}{r^2} - a_p\,,
\end{equation}  
which leads to the following effective gravitational mass of the Earth
\begin{equation}
G M_{\text{Earth}}^{eff} = G M_{Earth} + a_p r^2\,.
\end{equation}
The combined  LAGEOS and LLR measurements then requires
\begin{equation}
a_p (d_{\text{Earth-Moon}})^2 \leq 0.004 \frac{km^3}{s^2}
\end{equation}
while $a_p (d_{\text{Earth-Moon}})^2 = (0.129\pm 0.020) \frac{km^3}{s^2}$. Therefore the Earth-Moon system also refutes the covariant resolution  of Ref.  \cite{Saffari:2007zt} for the Pioneer anomaly. 

A lesson we should learn here is that ``phenomenologically good'' covariant corrections to the action which remain perturbative at the `close' vicinity of a spherical central mass seem to be  those whose predicted corrections to the space-time geometry had decreased if we would have decreased the central mass. The perturbative studies of $f(R_{ijkl}R^{ijkl})$ [this paper] and $f(R)$ \cite{Saffari:2007zt} show that this criterion is not  satisfied in general. It is interesting to systematically study  what kind of corrections meets this criterion. In so doing, perhaps, it is interesting to study $f(\frac{|R_{ijkl}R^{ijkl}|^3}{\nabla_iR_{jklm} \nabla^iR^{jklm}})$ that is suggested by the covariant resolution of the anomalous flat rotational curves of the spiral galaxies \cite{Exirifard:2008dy}.

\section{On the anomalous rotational velocity curves of the spiral galaxies}
\begin{figure}
   \epsfxsize=9cm
   \epsfbox{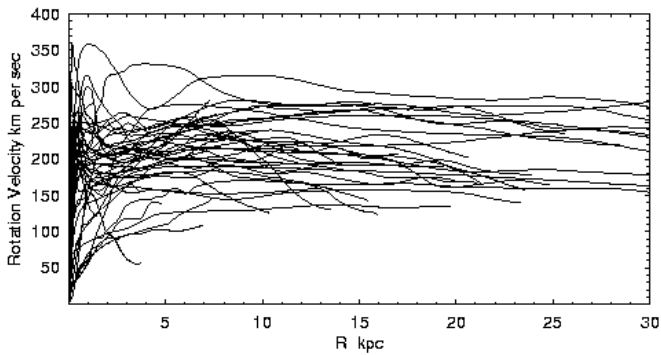}
   \caption{\label{fig2}\small \textit{Rotational curves of spiral galaxies obtained by combining CO data for the central regions, optical for disks, and HI for outer disk and halo \cite{Sofue:2000jx}.}}
\end{figure}

Eq. \eqref{ab-ln(r)} and \eqref{Veff} demonstrate that a correction in the form of $\Delta L=\alpha_{\frac{1}{2}}(R_{ijkl}R^{ijkl})^{\frac{1}{2}}$ leads to the following effective perturbative gravitational potential
\small
\begin{equation}\label{veff-found}
 V_{eff}(r)\, = \,-\frac{GM_g}{r}\,-\,2\,\sqrt{3}\,\alpha_{\frac{1}{2}}\,c^2\,\ln(\frac{r}{\text{r}_g})\,+\,O(\alpha^2_1)\,,
\end{equation}
\normalsize
around any spherical static distribution of matter of total inertial mass $M_g$. \eqref{veff-found} suggests that $\Delta L=\alpha_{\frac{1}{2}}(R_{ijkl}R^{ijkl})^{\frac{1}{2}}$ might have a chance to resolve the anomalous flat rotational velocity curves of the spiral galaxies without considering dark matter. Mathematically speaking, the current work does not approve or reject this suggestion, due to the following reasons:
\begin{enumerate}
 \item The anomalous velocity curves of the spiral galaxies occur at the boundaries of the spiral galaxies. The matter's distribution inside the presumed galaxy is disk-like and not spherical for the stars in the boundary of the spiral galaxies. \eqref{ab-ln(r)} and \eqref{Veff} are derived for a spherical distribution of matters.

\item The anomalous velocity curves of the spiral galaxies are not small deviation from what Newtonian gravity predicts. The exact solutions of the modified action might precede a possible resolution of the anomalous velocity curve.
\end{enumerate}
Despite the above obstacles we tend to examine whether a value of $\alpha_{\frac{1}{2}}$ compatible with \eqref{an-limit} has a chance to describe the flat rotational velocity curves of the spiral galaxies. In so doing, let us extrapolate \eqref{veff-found} toward the boundary of a typical spiral galaxy of mass $M_g=10^{12} M_\odot$. This generalization leads to the following relation for the velocity of the stars moving on a circular orbit around the center of the galaxy
\small
\begin{equation}\label{found-v}
v^2\,=\, \frac{G\,M_g}{r} + 2\, \sqrt{3}\, \alpha_{\frac{1}{2}}  c^2 + O(\alpha_{\frac{1}{2}}^2)\,,
\end{equation}
\normalsize
where $v$ stands for the velocity of the star around the center of the presumed galaxy. Examining the rotational curves of the spiral galaxies Fig.\ref{fig2}, we  see that the constant asymptotic velocity can be approximated by $200\pm50\frac{\text{km}}{\text{s}}$  in large scales. In these distances, the first term of \eqref{found-v} is small, thus \eqref{found-v} implies
\small
\begin{equation}\label{alpha-galaxy}
\alpha_{\frac{1}{2}}^g \,=\,( 13.61 \pm 6.39)\times 10^{-8} \,
\end{equation}
\normalsize
is needed to describe the constant velocity of the within the borders of a typical spiral galaxy. The high precision measurement around the Earth \eqref{an-limit}, however, requires
\small
\begin{equation}\label{alpha-half-limit}
\alpha_{\frac{1}{2}}^g \,\leq\, 6.67 \times 10^{-20} \,
\end{equation}
\normalsize
therefore, discards \eqref{alpha-galaxy} and implies that $\Delta L =(R_{ijkl}R^{ijkl})^{\frac{1}{2}}$ has no chance to describe the flat rotational velocity curves of the spiral galaxies.

It is worth noting that the Riemann scalar curvature in the Solar system and around the Earth satisfies
\small
\begin{equation}\label{R-condition}
\, \frac{7.53\times 10^{-71}}{\text{meter}^4}\,\leq\,
 R_{\mu\nu\eta\gamma}R^{\mu\nu\eta\gamma}\,,
\end{equation}
\normalsize
while in the regime where the anomalous rotational curvature of the spiral galaxies happens, it satisfies
\small
\begin{equation}\label{R-galaxy-condition}
\, \frac{ 10^{-104}}{\text{meter}^4}\,\lesssim   R_{\mu\nu\eta\gamma}R^{\mu\nu\eta\gamma}\, \lesssim
\, \frac{ 10^{-92}}{\text{meter}^4}\,,
\end{equation}
\normalsize
where a galaxy with a central mass at order $10^{12}M_{\odot}$ with the boundary of about $100$kpc is alleged. If we assume that a simple action in the form of a polynomial in terms of the Riemann tensor dictates the dynamics of the space-time in both of the regimes given by \eqref{R-condition} and \eqref{R-galaxy-condition} we observe that the value of $\alpha_{\frac{1}{2}}$ will be needed to describe the  anomalous flat rotational curve of the spiral galaxies \eqref{alpha-galaxy} is not in agreement with the accurate measurements of the space-time geometry around the Earth, as illustrated in Fig.\ref{fig1} as well. We, however, lack experimental justification or observational data supporting this assumption. Any functional of the Riemann tensor squared , $\Theta[{\cal R}^2]=\Theta[R_{\mu\nu\eta\gamma}R^{\mu\nu\eta\gamma}]$, which becomes sufficiently small for \eqref{R-condition} but constant for \eqref{R-galaxy-condition} can be utilized to suggest the following phenomenological action for gravity:
\small
\begin{equation}\label{gravity-ugly}
S\,=\,\int d^4 x\,\sqrt{-\det g} (R + \alpha_{\frac{1}{2}}^g ~\Theta[{\cal R}^2]~ (R_{\mu\nu\eta\gamma}R^{\mu\nu\eta\gamma})^{\frac{1}{2}})\,,
\end{equation}
\normalsize
which has a chance not only to be consistent with the Solar system's data but meets \eqref{alpha-galaxy} as well. The exact solutions of \eqref{alpha-galaxy} precedes reaching a concrete conclusion on the validity of the above suggestion. Addressing the exact solution of \eqref{alpha-galaxy} or a similar action in which $R_{\mu\nu\eta\gamma}R^{\mu\nu\eta\gamma}$ is replaced with the Gauss-Bonnet Lagrangian lays outside the scope of the current work.

\section{Conclusions}
We have approximated the space-time geometry in the Solar system by a Ricci flat geometry, a geometry of vanishing Ricci tensor. We have shown that a correction to the Einstein action would `non-trivially' perturbs the space-time geometry around a Ricci flat geometry in case the correction involves the Riemann tensor per se. After that we have considered  the simplest family of these corrections; the corrections which  are arbitrary functional of the Riemann tensor's squared:
\begin{subequations}%\label{RicciCorrections}
 \begin{align}
 L &= R + \epsilon \Delta L + O(\epsilon^2),\\
\Delta L &= L(R_{ijkl}R^{ijkl})\,,
 \end{align}
\end{subequations}
Then we have computed the corrections to the Schwarszchild black-hole in an asymptotically flat 4D geometry for a general $\Delta L$.

We have observed that $\Delta L_3=\alpha_{\frac{1}{3}}(R_{ijkl}R^{ijkl})^{\frac{1}{3}}$ gives rise to a constant anomalous acceleration for objects orbiting the Sun onward the Sun. This leaded us to the conclusion that $\alpha_{\frac{1}{3}}=(13.91\pm 2.11) \times 10^{-26}(\frac{1}{\text{meters}})^{\frac{2}{3}}$  would have covariantly resolved the Pioneer anomaly if this value of $\alpha_{\frac{1}{3}}$ had not contradicted with other observations.

We have shown that the experimental bounds on $\Delta L_3$ becomes stronger in case we examine the deformation of the space-time geometry around objects lighter than the Sun. We therefore have used the high precision measurements around the Earth (LAGEOS and Lunar Laser Ranging) and have obtained a strong constraint on the corrections in the form of $\Delta L(R_{ijkl}R^{ijkl})$ and in particular $\Delta L=\alpha_n(R_{ijkl}R^{ijkl})^n$. It is interesting that the high precision measurements around the Earth provide a strong constraint on the possible correction to the Einstein-Hilbert action.

The high precision measurements around the Earth requires $\alpha_{\frac{1}{3}}\leq6.12\times 10^{-29}(\frac{1}{\text{meters}})^{\frac{2}{3}}$, therefore, they  refute the  covariant resolution of the Pioneer anomaly. So the Pioneer anomaly can not be covariantly resolved within the general family of corrections we have considered. This supports the idea that the Pioneer anomaly is on board systematic or due to non-gravitational effects.

We also have noted that $\Delta L_2=\alpha_{\frac{1}{2}}(R_{ijkl}R^{ijkl})^{\frac{1}{2}}$ gives rise to an effective logarithmic gravitational potential. We have raised the question if  $\Delta L_2=\alpha_{\frac{1}{2}}(R_{ijkl}R^{ijkl})^{\frac{1}{2}}$ may be useful in describing the anomalous flat rotational velocity curves of the spiral galaxies, before having proved that a simple correction in the form of $\Delta L_2=\alpha_{\frac{1}{2}}(R_{ijkl}R^{ijkl})^{\frac{1}{2}}$ is not either in agreement with the high precision measurements around the Earth, or can not describe the flat rotational velocity curves of the spiral galaxies.

\section*{Acknowledgments}
 I thank  Mohammad Mehdi Sheikh-Jabbari and  Amir Hajian for fruitful  discussions. I  thank Mahmood Exirifard for his sharp comments on the English structure and writing  of the first version of the manuscript. I thank Peter Bender for informing me of the LAGEOS and LLR experiments by the criticism that he ruled out the prediction of the first version of this manuscript. I appreciate the comments and criticisms   made by anonymous referees of CQG.

\appendix
\section{Solar system tests \textit{do not} rule out 1/R gravity}
There exists a debate in the literature on the consistency of\footnote{This appendix is added as part of the response to a criticism made by the referee of CQG.} 
\begin{equation}\label{R-action}
S= \frac{1}{16 \pi G} \int d^4 x \sqrt{- \det g} (R - \frac{\mu^4}{R}) + \int d^4x \sqrt{-\det g } L_{m}
\end{equation}
for $\mu^{-1}\approx 10^{26} meters$ with the Solar system tests, for example look at \cite{Erickcek:2006vf,Chiba:2003ir,Kainulainen:2007bt,Olmo:2006eh,DeDeo:2007yn,Rajaraman:2003st,Multamaki:2006zb,Faraoni:2006hx,Ruggiero:2006qv,Allemandi:2005tg}. In particular ref. \cite{Erickcek:2006vf} claims  that properly matching the metric inside and outside the Sun rules out $\frac{1}{R}$ gravity. A sharp inspection of  \cite{Erickcek:2006vf}, however, reveals that \cite{Erickcek:2006vf} has not properly studied the equations. This was also noticed by \cite{DeDeo:2007yn}. In the following we would like to clarify why the conclusion of \cite{Erickcek:2006vf} is not right in addition to demonstrating the source of this wrong conclusion.

Contracting the equations of motion of \eqref{R-action} with the inverse of the metric yields
\begin{equation}\label{R-eq}
\Box \frac{\mu^4}{R^2} -\frac{R}{3} + \frac{\mu^4}{R} = \frac{8 \pi G T}{3}\,,
\end{equation}
where $T= g^{\mu\nu} T_{\mu\nu}$ and the speed of light is set one. Now let us define a new variable, $x$, through $R=- 8 \pi G x$. Rewriting \eqref{R-eq} in terms of $x$ and rearranging the terms yields
\begin{equation}\label{x-eq}
x = 1 - \frac{3 \mu^4}{(8 \pi G T)^2} (- \frac{1}{x} + \frac{1}{8 \pi G T} \Box \frac{1}{x^2} + O(\nabla T))\,.
\end{equation}
Note that the Einstein-Hilbert gravity holds $x=1$. We can obtain the order of magnitude of $T$ for the Sun by $T \approx \frac{3 M_\odot}{4 \pi R_\odot^3}$ where $R_\odot$ is the radius of the Sun. Knowing the order of magnitude of $T$, we can obtain the order of magnitude for deviation from $x=1$ in \eqref{x-eq}:
\begin{equation}
 \frac{3 \mu^4}{(8 \pi G T)^2} \approx  \frac{3 \mu^4}{(\frac{3 G M_\odot}{R_\odot^3 (\text{speed of light})^2})^2}= \frac{4 \mu^4 R_\odot^6}{3 r_\odot^2}\,,
\end{equation}
where $r_\odot = \frac{2 G M_\odot}{(\text{speed of light})^2}$ is the Schwarszchild radius associated to the mass of the Sun. Using $R_\odot = 1.39 \times 10^ 9 meters$, $r_\odot = 3 km$ and $\mu \approx 10^{-26} meters$, the order of magnitude of the deviation from $x=1$ in \eqref{x-eq} reads
\begin{equation}
 \frac{3 \mu^4}{(8 \pi G T)^2} \approx  10^{-55}.
\end{equation}
Note that the order of magnitude of the coffiecient in the front of $\Box \frac{1}{x^2}$ and similar terms in \eqref{x-eq} is $\frac{R_\odot^3}{r_\odot} \times \frac{3 \mu^4}{(8 \pi G T)^2}= 10^{-31} meters^2$. Therefore even the non-homogeneouty of the matter's distribution in the Sun does not produce a significant deviation from $x=1$ inside the Sun. 
Recalling that the Einstein-Hilbert gravity holds $x=1$ besides  extraordinarily  small deviation from $x=1$, we conclude that  the Einstein-Hilbert action quite-perfectly describes the physics in and outside the Sun for \eqref{R-action}.

Now let us inspect what leads  \cite{Erickcek:2006vf} to the wrong conclusion. Ref. \cite{Erickcek:2006vf} defines a function by
\begin{equation}
c = -\frac{1}{3} + \frac{\mu^4}{R^2}\,,
\end{equation}
which we refer to as the C-function. The authors  then take granted that the C-function encodes the deviation from the vacuum solution even outside the matter's distribution. This means that the authors fail to realize that  the C-function can be identically zero outside the matter's distribution.  In other words the deviation from the vacuum solution might  be encoded in other scalars rather the  Ricci scalar or equivalently the C-function.  They than rewrite \eqref{R-eq} in terms of the C-function
\begin{equation}\label{c-eq}
\Box c + \frac{\mu^2 c}{\sqrt{c+ \frac{1}{3}}} = \frac{8 \pi G  T}{3}, 
\end{equation}
before approximating it to
\begin{equation}\label{electro-eq}
\nabla^2 c = \frac{8 \pi G T}{3}\,.
\end{equation}
It then appears that the authors assume that the C-function and its derivatives are continuous on the surface of the Sun. But in the Enistein-Hilbert gravity  what remains continuous on the boundaries are the metric and its first derivatives. For example we know that the Ricci scalar or equivalently the C-function is not continuous on the boundary. Therefore instead of choosing C-function as what \cite{Erickcek:2006vf} has chosen, we must choose it in the following way
\begin{itemize}
\item Outside the Sun ($r>R_\odot$), $c=0$. Note that $c=0$ solves the equation outside the Sun.
\item Inside the Sun we must find a solution of \eqref{c-eq} that remains bounded inside the star.  
\end{itemize} 
The above choice means that within the electrostatic approximation to the equations - \eqref{electro-eq}-, the surface of the Sun effectively plays the role of a conducting surface accommodating some amount of  `charge' that completely cloaks the `charge' inside the Sun. Setting $c=0$ outside the Sun leads to the  Schawrszchild metric in the Solar system which is in agreement with observation. Therefore 1/R gravity is not ruled out at least due to reason addressed in  \cite{Erickcek:2006vf}.

\end{document}